\documentclass[a4paper,12pt,final]{article}
\usepackage{amsmath,amsfonts,amssymb,amsthm,amstext,eucal}

\def\bea{\begin{eqnarray}} 
\def\eea{\end{eqnarray}}
\def\beann{\begin{eqnarray*}} 
\def\eeann{\end{eqnarray*}}
\def\be{\begin{equation}} 
\def\ee{\end{equation}}
\def\ba{\begin{array}} 
\def\ea{\end{array}}
\def\ben{\begin{enumerate}} 
\def\een{\end{enumerate}}

\def\4{\tilde }
\def\5{\bar }  
\def\6{\partial } 
\def\7{\hat } 

\def\G{\Gamma}

\def\cL{{\cal L}}
\def\cE{{\cal E}}
\def\cF{{\cal F}}
\font\mybb=msbm10 at 10pt
\def\bb#1{\hbox{\mybb#1}}
\def\bR {\bb{R}}

\def\bE {\bb{E}}

\providecommand{\bysame}{\leavevmode\hbox to3em{\hrulefill}\thinspace}

\begin{document}


\begin{titlepage}
\vfill
\begin{flushright}
WIS/02/02-JAN-DPP\\
hep-th/0201061\\
\end{flushright}

\vfill

\begin{center}
\baselineskip=16pt
{\Large\bf D0-branes in flux 5-brane backgrounds}
\vskip 0.3cm
{\large {\sl }}
\vskip 10.mm
{\bf ~Joan Sim\'on }\\
\vskip 1cm
{\small
The Weizmann Institute of Science, Department of Particle Physics \\
Herzl Street 2, 76100 Rehovot, Israel \\
E-mail: jsimon@weizmann.ac.il}\\ 
\end{center}
\vfill
\par
\begin{center}
{\bf ABSTRACT}
\end{center}
\begin{quote}
The existence of supersymmetric D0-branes sitting at an arbitrary distance
from a flux 5-brane is proven. The physical picture in type IIA is consistent
with the Kaluza-Klein reduction origin of the flux 5-brane. An analysis of
the fluctuations around these vacua is performed and the backreaction of the
D0-branes is computed. Non-threshold D0-D2 bound states and similar 
stabilization mechanisms for D2-branes in such backgrounds are also briefly
discussed.
\end{quote}
\vfill
\vfill

\end{titlepage}


\section{Introduction}

There has recently been a lot of interest in the construction and 
classification of supersymmetric fluxbranes in string theory, either
from a supergravity perspective ~\cite{gutperle,fluxsugra} or from a 
conformal field theory (CFT) one ~\cite{fluxcft,russo}. It is natural to
ask about which D-branes, NS-branes and other dynamical objects in string 
theory exist in this new sector of the theory ~\cite{paper2}.
Some results in this direction were obtained using CFT and/or world-volume
techniques in ~\cite{branefluxcft}-~\cite{note}. In this note, we shall use
low energy field theory descriptions both in the open and closed string
sectors to answer this question, and for simplicity, we shall concentrate
on D0-branes in flux 5-branes (F5-branes) ~\cite{gutperle}. 

A F5-brane is a type IIA configuration preserving one half of the spacetime 
supersymmetry (see appendix) with metric

\begin{multline}
     g = \Lambda^{1/2}\left\{ds^2(\bE^{1,5}) + dr^2 + r^2(d\theta^2
     +\sin^2 2\theta d\psi^2)\right\} \\
     + r^2\Lambda^{-1/2}\left\{d\varphi+\cos 2\theta d\psi\right\}^2~,
  \label{f5brane}
\end{multline}
and nontrivial dilaton and Ramond-Ramond (RR) 1-form 
\begin{gather}
  \Phi - \Phi_0 = \frac{3}{4}\log\Lambda \\
  C_{(1)} = (\beta r)r\Lambda^{-1}(d\varphi+\cos 2\theta d\psi)\,,
 \label{RR}
\end{gather}
where the scalar function $\Lambda$ is defined by $\Lambda=1+(\beta r)^2$.
The above configuration is obtained by Kaluza--Klein reduction of eleven
dimensional Minkowski spacetime  along the
orbits of the Killing vector 
\[
  \xi_{\text{F5}}=\partial_{x^\sharp} + \beta\left(x^6\partial_7 - 
  x^7\partial_6 + x^8\partial_9 - x^9\partial_8\right)~,
\]
where $x^\sharp=R\,\chi$ stands for the eleventh compact dimension
and $\chi$ for some canonically normalised angular variable. Equivalently,
the F5-brane is obtained in the limit \footnote{The author would like
to thank H. Robins for discussions on this point.}
\begin{equation}
  R\to 0 \qquad , \qquad \beta~~\text{fixed}~.
 \label{limit}
\end{equation}
Notice that whenever $\beta r_0 \sim g_s^{-2/3}$,
the type IIA description of the geometry is no longer reliable, and one
should replace it by its M-theory lift. As explained in ~\cite{gutperle}, one 
can associate some notion of ``charge'' to this object by computing the
integral $\int F_{(2)}\wedge F_{(2)}$, on the full $\bR^4$ transverse space 
to the F5-brane, $F_{(2)}$ being the corresponding RR 2-form field strength. 
This transverse $\bR^4$ space is parametrised by
\begin{equation}
  \begin{aligned}[m]
    x^6 + ix^7 &= r\cos\theta e^{i(\varphi+\psi)} \\
    x^8 + ix^9 &= r\sin\theta e^{i(\varphi-\psi)} ~.
  \end{aligned}
\end{equation} 
Thus, $r$ stands for the radial distance to the F5-brane, $\theta$ measures 
the angle between the 67-plane and the 89-plane, whereas $\varphi\pm
\psi$ are the polar angles (phases) in the corresponding 2-planes. The charge
(proportional to $(\beta^2)^{-1}$) is thus kept fixed in the Kaluza--Klein
reduction \eqref{limit}.

Assuming the existence of D0-branes in this background at weak coupling, and
taking a strong coupling limit, the system should be described by an M-wave
~\cite{hull} propagating in a locally flat spacetime satisfying topologically
nontrivial identification conditions \footnote{Similar remarks were already
mentioned and used in the context of M5-branes in the appendix of 
~\cite{russo}.}. It is thus expected that the Kaluza--Klein reduction
of an M-wave configuration along the orbits of the Killing vector 
$\xi_{\text{F5}}$ should give rise to a type IIA configuration describing
a composite system of D0-branes and F5-branes. One of the purposes of this 
note is to check this interpretation.

In order to address these considerations, one can study the dynamics at low
energy of a single D0-brane in such an F5-brane background, by a probe 
analysis. If a D0-brane of tension (mass) $\text{T}_{\text{D}0}=
\frac{1}{l_s}e^{-\Phi_0}$ sits in a F5-brane background at a distance $r_0$, 
the potential felt by the probe
\begin{equation}
  V=T_{\text{D}0}\left[1+3\left(\beta r\right)^2\Lambda_0^{-1}\right]^{1/2}~,
 \label{potentiald0}
\end{equation}
depends on the distance. In \eqref{potentiald0}, $\Lambda_0$ was defined as
$\Lambda(r_0)$. It is only at $r_0=0$, the minimum of the potential
\eqref{potentiald0}, that the probe feels no force. Equivalently, the 
classical equations of motion for a massive particle propagating in a F5-brane
background with no angular velocity are only satisfied when motion is
constrained to the hypersurface defining the F5-brane. 

It is a natural question whether there exists some physical mechanism
by which the D0-brane remains {\sl static}\footnote{In this note, by static, 
we shall refer to a configuration having vanishing angular momentum.} 
at an arbitrary distance $r_0$ from the F5-brane. The answer to this question
lies on the geometry of the background. Notice that even if the 
angular velocity of the D0-brane vanishes, the RR 1-form induces some angular 
momentum through the Wess-Zumino coupling, yielding a non-static 
configuration. This is the reason why \eqref{potentiald0} is not equal to 
minus the lagrangian density, but includes the contribution from such 
non-vanishing momenta. This argument raises the possibility of giving some 
angular velocity to the D0-brane such that the total angular momentum 
vanishes, yielding an static configuration of constant energy 
(independent of $r_0$). This expectation will be confirmed both for
a single D0-brane in type IIA and a massless particle in eleven dimensions
(which is the lift to M-theory of the type IIA configuration).

The plan of the paper is as follows.
In section 2, we shall study the propagation of a massless 
eleven dimensional particle in a flat eleven dimensional geometry satisfying
some nontrivial identification conditions, by using an adapted coordinate
system to the action of $\xi_{\text{F5}}$. In section 3, we shall move to the 
D0-brane setting and prove, by construction, the existence of the 
forementioned {\em static} configurations of finite energy. 
Evidence is given in favour of their stability by an analysis of the 
fluctuations around these vacua and due to the existence of a bound from 
below on their energy. In section 4, the supergravity type IIA configuration 
taking into the account the backreaction of a bunch of D0-branes is found. 
It is then shown that an additional D0-brane probe having the same angular 
velocity as in our previous vacua can be added at an arbitrary distance, 
thus showing its BPS character, and confirming the physical interpretation of 
these supergravity backgrounds. The same physical effect is used to
construct D0-D2 bound states in section 5 and to (meta)-stabilize curved 
D2-branes in F5-brane backgrounds in section 6.

\section{Eleven dimensional massless particle}

Let us study the propagation of an eleven dimensional massless particle,
\begin{equation}
  S=\int d\tau\,\cL = \int d\tau\left[\dot{x}^MP_M -
  \frac{1}{2}vh^{MN}P_MP_N\right] ~,
 \label{massless}
\end{equation}
in the eleven dimensional geometry 
\begin{multline}
  h = ds^2(\bE^{(1,5)}) + dr^2 + r^2(d\theta^2 + \sin^2 2\theta
  d\psi^2) + \Lambda R^2 d\chi^2 \\
  +  2\beta r^2 R\,d\chi (d\varphi+\cos 2\theta d\psi) + 
  r^2 (d\varphi + \cos 2\theta d\psi)^2~. 
 \label{adapted}
\end{multline}
In \eqref{massless}, $P_M$ $M=0,1,\dots ,\sharp$ should be thought of some
set of auxiliary fields, actually, the conjugate momentum variables to
the spacetime coordinates $x^M$; whereas $v$ ensures the mass-shell condition.
The eleven dimensional metric \eqref{adapted} was already written in adapted 
coordinates, where $\xi_{\text{F5}}=\partial_{x^\sharp}$. Thus, by 
Kaluza--Klein reduction  along the $x_\sharp=R\chi$ direction, the full 
F5-brane background is obtained.

Its classical equations of motion are given by,
\begin{equation}
  \begin{aligned}[m]
     h^{MN}P_MP_N &= 0 \\
     P_M &= h_{MN}\dot{x}^N\,v^{-1} \\
     \frac{d\,P_M}{d\tau} &= \frac{\6\cL}{\6 x^M}
  \end{aligned}
  \label{motionmassless}
\end{equation}
where $v$ can always be set to one (using worldline reparametrisations)
and the inverse of the background metric $h^{MN}$ is
\begin{equation*}
  h^{-1} = 
  \begin{pmatrix}
    \eta^{\mu\nu} & 0 & 0 & 0 \\
    0 & 1 & 0 & 0 \\
    0 & 0 & r^{-2} & 0 \\
    0 & 0 & 0 & \tilde{h}^{-1}
  \end{pmatrix}
  \qquad , \qquad \mu\,\nu=0,1,\dots ,5
\end{equation*}
\begin{equation*}
  \tilde{h}^{-1} =
  \begin{pmatrix}
    R^{-2} & -\beta\,R^{-1} & 0 \\
    -\beta\,R^{-1} & \Delta\left(1+\beta^2 r^2 \sin^2 2\theta\right)  
    & -\Delta\cos 2\theta \\
    0 & -\Delta\cos 2\theta & \Delta
  \end{pmatrix}
\end{equation*}
where the scalar function $\Delta=(r\sin 2\theta)^{-2}$ was introduced and
$\tilde{h}^{-1}$ is written in the basis $\{\chi\,,\varphi\,,\psi\}$.

It will be assumed that all momenta is vanishing except for $P_\chi$, 
$P_\varphi$
and $P_\psi$, which are conserved quantities,  this being consistent with 
(\ref{motionmassless}) and equivalent to setting $\dot{r}=\dot{\theta}=
\dot{x}^i=0$ $i=1,\dots ,5$. The energy of the configuration is given by the 
mass-shell condition
\begin{equation}
  E^2 = \left(\frac{P_\chi}{R}+ \beta P_\varphi\right)^2 +
  \frac{1}{r_0^2\sin^2 2\theta_0}\left(P_\varphi-\cos 2\theta_0 P_\psi\right)^2
  +(P_\psi r_0^{-1})^2
\end{equation}
and depends on the constant parameters $r_0,\theta_0$. Requiring the energy
to be minimized enforces an stationary requirement in the $\varphi$, $\psi$
angular directions 
\[
P_\varphi=P_\psi =0\,,
\]
where the solution $r_0\to\infty$ was excluded, since we are interested
in finite energy configurations. 

Using (\ref{motionmassless}), we can solve the stationary conditions in terms
of the velocities
\begin{equation}
  \begin{aligned}[m]
     r^2[\dot{\varphi} + \cos 2\theta\dot{\psi} + \beta\,R\dot{\chi}] &= 0 \\
     r^2[\dot{\psi} + \cos 2\theta\dot{\varphi} 
     + \beta\,R\cos 2\theta\dot{\chi}] 
     &= 0 \\
     \Lambda R^2 \dot{\chi} + \beta\,R r^2(\dot{\varphi} 
     + \cos 2\theta\dot{\psi})
     &= P_\chi ~,
  \end{aligned}
\end{equation}
allowing three different solutions
\begin{itemize}
  \item
    $r_0=0$, corresponding to a massless particle constrained to 
    $\bR^5\times S^1$.
  \item
    $\dot{\psi}=0$, $\dot{\varphi}=-\beta R^{-1}\,P_\chi
    =-\beta P_\sharp$, $\forall\,r_0$.
  \item
    $\theta_0=0(\pi/2)$, $\dot{\varphi}+(-)\dot{\psi}=
    -\beta P_\sharp$, $\forall\,r_0$, corresponding to a massless particle 
    moving in the 67-plane (89-plane).
\end{itemize}
In all cases, the energy of the particle equals its velocity
\[
E=\dot{x}_\sharp ~,
\]
along the compact direction, since $P_\chi=R^2\dot{\chi}=R\dot{x}_\sharp$,
thus suggesting the latter configurations do describe a massless particle
propagating in the compact direction $x^\sharp$. This can explicitly
be shown by mapping the trajectories in both the adapted coordinate
system and the one in which the metric is manifestly flat. The latter
is governed by
\[
  \phi = \varphi + \beta\,x^\sharp~.
\]
Thus, whenever $\dot{\psi}=0$, there is no angular velocity in $\phi$
$\left(\dot{\phi} = 0\right)$; whenever $\theta_0 = 0 (\pi/2)$, motion
is constrained to a plane, thus the real physical phase variable is
given by $\phi+\cos 2\theta_0\psi$, which again has vanishing angular
velocity; for $r_0=0$, no motion is allowed in the directions under
consideration. Thus, indeed, all the found trajectories do correspond
to propagation in the compact direction, as the energy computation
was suggesting to us. 

All configurations described in this section preserve one quarter of the
spacetime supersymmetry. This is obvious in the coordinate system in which 
the background is manifestly flat. Indeed, in that frame, the background 
preserves the supersymmetries satisfying $\G_{6789}\varepsilon=\varepsilon$,
whereas the propagating massless particle the ones obeying $\G_0\G_\sharp
\varepsilon=\varepsilon$. Since $[\G_{6789}\,,\G_0\G_\sharp]=
\text{tr}\,\G_{6789}\G_0\G_\sharp = 0$, the configuration preserves
eight supercharges.

\section{Static D0-brane in a F5-brane background}

Once the motion of massless eleven dimensional particle in the adapted
coordinate system has been worked out, the extension to the D0-brane
is rather natural. One should look for an angular velocity such that the 
full angular momentum of the massive particle vanishes. We shall consider
the most general ansatz consistent with this picture, in which
$x^0 = \tau$ (gauge choice), the coordinates parametrising the
F5-branes worldspace are set to a constant value $(x^i_0)$, the radial
distance to the F5-brane $(r_0)$ and $\theta=\theta_0$ are constants
and the angular coordinates that do couple with the RR 1-form (\ref{RR})
are assumed to be
\begin{equation}
  \begin{aligned}[m]
    \varphi &= \omega_0\tau + \varphi_0 \\
    \psi &= v_0\tau + \psi_0~.
  \end{aligned}
 \label{d0ansatz}
\end{equation} 
The action describing the dynamics of this system at low energies and
weak string coupling constant is given by 
\begin{equation}
  \begin{aligned}[m]
    S_{\text{D}0} &= -T_{\text{D}0}
    \int d\tau\Lambda_0^{-1}\sqrt{\Lambda_0-r_0^2
    (\omega_0+v_0\cos 2\theta_0)^2
    -\Lambda_0 (r_0 v_0\sin 2\theta_0)^2} \\
    & + T_{\text{D}0}\int d\tau \Lambda_0^{-1}\beta 
    r_0^2(\omega_0+v_0\cos 2\theta_0)~.
  \end{aligned}
 \label{d0action}
\end{equation}
The non-trivial components of the angular momentum can be derived
from (\ref{d0action}) by standard methods. In particular, computing
$\partial\cL/\partial\omega_0$ one obtains
\begin{equation}
  P_\varphi = T_{\text{D}0}r_0^2 \Lambda_0^{-1}\left[\beta +
  \frac{l_0}{\sqrt{\Lambda_0-(r_0 l_0)^2-\Lambda_0 (r_0 v_0\sin 2\theta_0)^2}}
  \right]
 \label{moment1}
\end{equation}
and from $\partial\cL/\partial v_0$
\begin{equation}
  P_\psi = T_{\text{D}0}r_0^2 \Lambda_0^{-1}\left[\beta\cos 2\theta_0 +
  \frac{l_0\cos 2\theta_0 + \Lambda_0 r_0^2\sin^2 2\theta_0 v_0}
  {\sqrt{\Lambda_0- (r_0 l_0)^2-\Lambda_0 (r_0 v_0\sin 2\theta_0)^2}}\right]~,
 \label{moment2}
\end{equation}
where the constant parameter, $l_0=\omega_0+v_0\cos 2\theta_0$, has been
introduced.

When one requires both equation (\ref{moment1}) and equation (\ref{moment2}) 
to vanish, one finds three different solutions to the corresponding system of 
equations :
\begin{itemize}
  \item
    $r_0=0$, corresponding to a D0-brane restricted to move on the F5-brane.
  \item
    $\theta_0=0,\pi/2$, $\omega_0\pm v_0 = -\beta$ $\forall$ $r_0$, 
    corresponding to a D0-brane moving either in the 67 plane 
    $(\theta_0=0)$ or in the 89 plane $(\theta_0=\pi/2)$ with total angular 
    velocity $-\beta$, in both cases.
  \item
    $v_0=0$, $\omega_0 = -\beta$ $\forall\,\theta_0\,,r_0$.
\end{itemize}
It is reaffirming to get exactly the same conditions we derived before.
Since, by construction, the angular momentum vanishes, the energy of the
configuration equals minus the lagrangian density of the system 
$\left(E=-{\cal L}\right)$, that is, it equals the tension (mass) of the 
D0-brane. Thus, there exist configurations of {\sl static} D0-branes sitting 
at any distance $r_0$ from the F5-branes, with non-vanishing angular velocity 
and finite energy equal to its mass. This interpretation can also be derived 
from a pure hamiltonian analysis of the system. Using the general formalism 
developed in ~\cite{ericpaul}, the energy density can be computed from the 
mass-shell condition
\begin{equation} 
  \4P^2 + T_{\text{D}0}^2\Lambda_0^{-3/2}=0 ~,
\end{equation}
where $\4P^2 = g^{MN}(P_M+T_{\text{D}0}C_M)(P_N+T_{\text{D}0}C_N)$, 
$g^{MN}$ being the inverse of the metric (\ref{f5brane}) and $C_M$ the 
components of the RR 1-form (\ref{RR}). Notice that by requiring the spacelike 
components of the momentum $(P_m=0 \,\,m=1,\dots ,9)$ to vanish as 
corresponds to a {\sl static} configuration, the energy $(P_0=E)$ equals the 
tension (mass) of the D0-brane.

Given a bosonic configuration on the worldvolume of any brane, one way
to check whether it preserves some supersymmetry is by analysing whether
it satisfies the kappa symmetry preserving condition ~\cite{kappasusy}
\begin{equation}
  \Gamma_\kappa\varepsilon=\varepsilon 
 \label{kappad0}
\end{equation}
where $\varepsilon$ is the Killing spinor of the corresponding background
geometry and $\Gamma_\kappa$ is the usual field and background dependent 
matrix encoding the kappa symmetry transformations in
D-branes ~\cite{cederwall}. Equation \eqref{kappad0} reduces, in
this particular case, to
\begin{equation}
  \left\{\Gamma_{\underline{0}}+ \Lambda_0^{-1/2}r_0 l_0
  \Gamma_{\underline{\varphi}} + \Lambda_0^{1/2}r_0v_0\sin 2\theta_0
  \Gamma_{\underline{\psi}}\right\}
  \Gamma_{11}\varepsilon=\varepsilon\,.
 \label{kd0a}
\end{equation}
Thus, for $r_0=0$, equation \eqref{kd0a} reduces to
\[
\Gamma_{\underline{0}}\Gamma_{11}\varepsilon=\varepsilon \quad (r_0=0) ~,
\]
whereas in the remaining two static configurations, \eqref{kappad0} reduces to
\[
  \left\{\Gamma_{\underline{0}} - \Lambda_0^{-1/2}
  (\beta r_0)\Gamma_{\underline{\varphi}}\right\}
  \Gamma_{11}\varepsilon=\varepsilon \quad \left(l_0 = -\beta\right) ~,
\]
which corresponds to the supersymmetry projection condition of a boosted 
D0-brane. As it is argued in the appendix, the point dependence of the
Killing spinors is of the form $\varepsilon=\tilde{M}[x;\Gamma]\varepsilon_0$,
where $\varepsilon_0$ stands for a constant spinor and 
$\tilde{M}[x;\Gamma]$ is some matrix, depending on the point, built from flat 
gamma matrices. Under such conditions, the analysis of \eqref{kd0a} is not 
straightforward, as has already been pointed out in the past in different 
backgrounds ~\cite{ntkappa}. 

An explicit computation involving the exact form for the Killing spinors
should prove that our static configurations preserve one fourth of the 
spacetime supersymmetry. To give further evidence to the last statement,
we shall first study the stability of these configurations by an analysis of
the fluctuations around them and the existence of a bound from below
for the energy, and shall postpone the discussion on the BPS character
of these configurations to the next section. Let us parametrise these arbitrary
fluctuations by
\begin{equation}
  \begin{aligned}[m] 
    x^i &= x^i_0 + \eta\,\delta x^i(\tau) \\
    r &= r_0 + \eta\,\delta r(\tau) \qquad , \qquad
    \theta=\theta_0 + \eta \,\delta\theta (\tau) \\
    \varphi &= \omega_0\tau + \varphi_0 + \eta\,\delta\varphi (\tau) \\
    \psi &= v_0\tau + \psi_0 + \eta\,\delta\psi (\tau)~.
  \end{aligned}
\end{equation}
where $\eta$ will be the parameter counting the different orders in the 
expansion of the action around the vacuum configuration, 
$S_{\text{D}0}=S_{(0)} + \eta\,S_{(1)} + \eta^2\,S_{(2)} + {\cal O}(\eta^3)$. 

Let us study the different contributions to the expansion of the action
$S_{\text{D}0}$ by splitting the latter into a Dirac-Born-Infeld (DBI) part 
and a Wess-Zumino (WZ), $S_{\text{D}0}=S_{\text{DBI}}+S_{\text{WZ}}$. The 
contribution from the WZ term is given by
\begin{equation}
  \begin{aligned}[m]
    S_{\text{WZ}} &= T_{\text{D}0}\int d\tau 
    \left\{\frac{\beta r_0^2}{\Lambda_0}l_0 +
    \eta\left[\frac{\beta r_0^2}{\Lambda_0}(\delta\dot{\varphi}
    +\delta\dot{\psi}\cos 2\theta_0) 
    + 2 \frac{\beta r_0^2}{\Lambda_0}l_0 \delta r\right] \right.\\
    & + \left. \eta^2\left[-2\frac{\beta r_0^2}{\Lambda_0}
    (v_0\cos 2\theta_0\delta\theta^2 +
    \sin 2\theta_0 \delta\theta \delta\dot{\psi}) \right.\right. \\ 
    & \left. \left. + \frac{\beta}{\Lambda_0^3}
    l_0(1-3(\beta r_0)^2)\delta r^2 + 2\frac{\beta r_0}{\Lambda_0^2} \delta r
    (\delta\dot{\varphi}+\cos 2\theta_0\delta\dot{\psi})\right]\right\}~,
  \end{aligned}
 \label{wzfluc}
\end{equation}
where we already took into account that $r_0 v_0\sin 2\theta_0=0$ for any of 
the three classical solutions found previously.

In order to write the expansion of the DBI term, we shall distinguish
between the $r_0=0$ classical solution and the ones involving an arbitrary
$r_0$. Notice that for $r_0=0$, the contribution from the WZ term 
(\ref{wzfluc}) vanishes and one is left with
\begin{equation}
  \begin{aligned}[m]
    S_{\text{D}0} & = -T_{\text{D}0}\int d\tau 
    + \eta^2 T_{\text{D}0}\int d\tau \left\{
    \frac{1}{2}\delta\dot{x}^i\delta\dot{x}^j\delta_{ij} +
    \frac{1}{2}\delta\dot{r}^2 + \frac{1}{2}(\beta)^2\delta r^2\right\} \\
    & + {\cal O}(\eta^3)~.
  \end{aligned}
 \label{case1}
\end{equation}
The zeroth order contribution in the $\eta$ parameter reproduces the energy
of the vacuum configuration (up to a sign), as it should, whereas the first
order contribution vanishes, in agreement with the fact that the classical
configuration which we are expanding around solves the classical 
equations of motion. The second order contribution describes the dynamics
of five massless scalar fields (the fluctuations along the five spacelike
directions along the F5-brane) plus a radion $\delta r$, whose mass term
has the wrong sign.

Let us now concentrate on the remaining two cases $(r_0\neq 0)$.
The contribution from the DBI term can be summarized as
\begin{equation}
  \begin{aligned}[m]
    S_{\text{DBI}}&= -T_{\text{D}0}\int d\tau \Lambda_0^{-1}
    \left\{1+\eta\left(\frac{1}{2}\cL_1 -x_1\right) \right. \\
    & + \left. \eta^2\left(\frac{1}{2}(\cL_2 - \frac{1}{4}\cL_1^2) +
    x_1^2 -x_2 -\frac{1}{2}x_1\cL_1\right)\right\}~,
  \end{aligned}
 \label{dbifluc}
\end{equation}
where we have defined
\begin{equation}
 \begin{aligned}[m]
   x_1 &= 2 \frac{\beta^2 r_0}{\Lambda_0}\delta r \qquad , \qquad
   x_2 = \frac{\beta^2}{\Lambda_0}\delta r^2 \\
   \cL_1 &= 2(\beta r_0)r_0(\delta\dot{\varphi}
   +\cos 2\theta_0 \delta\dot{\psi})\\
   \cL_2 &= 4(\beta r_0)\delta r(\delta\dot{\varphi}+\cos 2\theta_0 
   \delta\dot{\psi}) -\Lambda_0\delta\dot{x}^i\delta\dot{x}^j\delta_{ij} 
   - \Lambda_0\delta\dot{r}^2 -\Lambda_0 r_0^2 \delta\dot{\theta}^2 \\
   &- 4(\beta r_0)r_0(v_0\cos 2\theta_0
   \delta\theta^2 + \delta\dot{\psi}\sin 2\theta_0 \delta\theta) -
   r_0^2(\delta\dot{\varphi}+\cos 2\theta_0 \delta\dot{\psi})^2 \\
   &-\Lambda_0r_0^2\sin^2 2\theta_0 \delta\dot{\psi}^2 -\Lambda_0 v_0^2
   (4r_0^2 \cos 4\theta_0 \delta\theta^2 + \sin^2 2\theta_0\delta r^2)~.
  \end{aligned}
 \label{inter}
\end{equation}
Joining the contributions from (\ref{wzfluc}) and (\ref{dbifluc}), we recover
the zeroth order contribution corresponding to the vacuum energy.
At linear order in the expansion parameter $\eta$, the action
\begin{equation}
  S_{(1)} = T_{\text{D}0}\int d\tau \Lambda_0^{-1}\left(x_1-
  2\frac{\beta^2 r_0}{\Lambda_0}\delta r\right)
\end{equation}
vanishes, as can be seen by the definition of $x_1$ in (\ref{inter}). 
Finally, at second order, the action reads as follows
\begin{equation}
  \begin{aligned}[m]
    S_{(2)} &= T_{\text{D}0}\int d\tau \left\{\frac{1}{2}\delta\dot{x}^i
    \delta\dot{x}^j\delta_{ij} + \frac{1}{2}\delta\dot{r}^2 +
    \frac{1}{2}r_0^2 \delta\dot{\theta}^2 + \frac{1}{2}r_0^2
    (\delta\dot{\varphi}+\cos 2\theta_0 \delta\dot{\psi})^2 \right. \\
    & + \left. \frac{1}{2} r_0^2\sin^2 2\theta_0 \delta\dot{\psi}^2 +
    2r_0^2 v_0^2 \cos 4\theta_0 \delta\theta^2 + 
    \frac{1}{2}v_0^2 \sin^2 2\theta_0 \delta r^2 \right\}~.
  \end{aligned}
\label{actionfluc}
\end{equation} 
The above action describes a free scalar field theory in which
only two mass like terms appear with the wrong sign. Notice that 
$v_0\sin 2\theta_0$ is always
vanishing for the vacua we are discussing, thus radial fluctuations
are both classically stable and massless. Furthermore, when $v_0=0$, the mass
term for $\delta\theta$ fluctuations vanish, thus we conclude that
the static classical configuration defined by $v_0=0$
and $\omega_0=-\beta$ is a stable configuration in which all fluctuations 
$\{\delta x^i\,,\delta r\,,\delta\theta\,, r_0\left(\delta\varphi +
\cos 2\theta_0\delta\psi\right)\,, r_0\sin 2\theta_0\delta\psi\}$ are 
massless, as it should be for a D0-brane with no constraints on its motion.

Finally, for the classical configuration confined either
to the 67-plane $(\theta_0=0)$ or to the 89-plane $(\theta_0=\pi/2)$
the wrong sign mass like term for the fluctuation $\delta\theta$ remains.
Notice that in this case, $r_0\sin 2\theta_0\delta\psi$ is also non-dynamical 
since the only physical degree of freedom corresponds to fluctuations of the 
phase $r_0\left(\delta\varphi\pm\delta\psi\right)$ in the plane where the 
classical motion takes place. 

Despite the apparent tachyonic behaviour of the radion $\delta r$ for the 
$r_0=0$ vacuum and $\delta\theta$ for the $\theta_0=0(\pi/2)$ vacua, one 
should check that the energy of these fluctuations is negative, to consider
them as real tachyonic modes. Actually, such a possibility is not possible.
Indeed, by solving the mass-shell condition $\tilde{P}^2 + T_{\text{D}0}
\Lambda^{-3/2}=0$ (in general), one can express the energy of the system
as
\begin{multline}
  E^2 = \left(T_{\text{D}0}+\beta P_\varphi\right)^2 + \left(\frac{P_\psi}
  {r}\right)^2 + \frac{1}{r^2\sin^2 2\theta}\left(P_\varphi -
  \cos 2\theta P_\psi\right)^2 \\
  + P_r^2 + \left(\frac{P_\theta}{r}\right)^2 + \sum_{i=1}^5 
  \left(P_i\right)^2\,.
 \label{bound}
\end{multline}
Thus, energy is bounded from below by $T_{\text{D}0}$, a bound which is
saturated when the configuration is static $\left(P_\varphi=P_\psi=P_r=
P_\theta=P_i=0\right.$ $\left.\forall\,i\right)$. Due to the positivity of 
all the terms in \eqref{bound}, any fluctuation around our vacua would give 
rise to a positive energy contribution, thus showing the stability of all 
configurations discussed before.

\section{Background reaction}

If the above configuration is BPS, one could consider a set of N D0-branes
sitting on the same point in the presence of a F5-brane. For large enough
N, such a configuration should admit a reliable supergravity description.
Once this background is known, one could probe it with an additional D0-brane
and the BPS feature should manifest itself with the existence of classical
configurations having the same features as the ones described
in the previous section, and allowing the additional D0-brane to sit at
an arbitrary distance not only of the F5-brane but also of the remaining
D0-branes being described by the background. In the following, we shall
first construct the closed string description of N D0-branes in the presence
of a F5-brane, and afterwards, it will be checked that indeed an additional
D0-brane can sit at an arbitrary distance when having the correct
angular velocity.

\subsection{Supergravity background}

The easiest way to look for the type IIA configuration we are interested
in is to take the strong coupling limit of the theory $(R\to \infty)$,
and look for the corresponding configuration in M-theory \footnote{This remark
was already mentioned in the appendix of \cite{russo}.}. This eleven
dimensional configuration should be an M-wave propagating in the compact 
direction, plus some topologically non-trivial conditions describing the 
presence of the F5-brane. Thus, locally, the metric should look like the 
one of an M-wave ~\cite{hull}
\begin{equation}
  h = \left(U-2\right)(dx^0)^2 + U\,(dx^\sharp)^2 - 2\left(U-1\right)
  dx^0dx^\sharp + ds^2(\bE^9)~,
\end{equation}
where $x^\sharp$ stands for the direction of propagation and $U=U(\7r)$
is an harmonic function 
\[
  U=U(\7r)= 1+ \frac{k}{\7r^7}~,
\]
on $\bE^9$. Notice that $\7r^2 = r^2 + x^ix^j\delta_{ij}$, where $r$ is
the radial coordinate transverse to the F5-brane and $x^i$ $(i=1\,\dots ,5)$
parametrise its worldspace. The configuration preserves one fourth
of the supersymmetry. This is because the M-wave configuration is already
a one-half BPS object, whereas the nontrivial identifications that
characterise the global properties of this configuration break another
half.

If one Kaluza-Klein reduces the above configuration along the orbits
of the Killing vector $\xi_{\text{F5}}=\partial_{x^\sharp} 
+ \beta\left(x^6\6_7-x^7\6_6 + x^8\6_9 - x^9\6_8\right)$ \footnote{This
is just a particular case of a more general formulation discussed in
~\cite{paper2}.}, one obtains a type IIA configuration with metric
\begin{multline}
  g = -\Delta^{-1/2}\,U^{-1/2}\left(U-1\right)^2(dx^0)^2 
  + \Delta^{-1/2}\,U^{1/2} r^2\left(d\varphi+\cos 2\theta d\psi\right)^2 \\
  + 2\beta r^2
  \Delta^{-1/2}\,U^{-1/2}\left(U-1\right)dx^0\,\left(d\varphi+\cos 2\theta
  d\psi\right) + \Delta^{1/2}\left\{U^{1/2}\left(U-2\right)(dx^0)^2 \right. \\
  \left. + U^{1/2}ds^2(\bE^5)
  + U^{1/2}\left(dr^2 + r^2\left(d\theta^2+\sin^2 2\theta d\psi^2\right)
  \right)\right\}
 \label{d0f5}
\end{multline}
RR 1-form potential $C_{(1)}$
\begin{equation}
  C_{(1)} = \Delta^{-1}\left\{-\left(1-U^{-1}\right)dx^0 + U^{-1}\,
  \beta r^2 \left(d\varphi+\cos 2\theta d\psi\right)\right\}~,
\end{equation}
and a non-vanishing dilaton
\begin{equation}
  \phi = \frac{3}{4}\log\left(U\cdot\Delta\right)~,
\end{equation}
where the scalar function $\Delta$ is defined as
\[
\Delta = 1 + U^{-1}\,\beta^2 r^2 ~.
\]

Notice that this is a type IIA background that correctly reproduces
the F5-brane limit, the one in which the charge of the D0-brane is sent to zero
$(k\to 0)$, and the D0-brane limit, in which the charge of the F5-brane
is sent to zero $(\beta\to 0)$. In both limits, there is an enhancement
of supersymmetry. Furthermore, the ten dimensional metric \eqref{d0f5} 
takes into account the backreaction of the boosted
D0-branes by having non-vanishing $g_{0\varphi}$ and $g_{0\psi}$ components.

\subsection{Probe computation}

We shall now explicitly show that exactly the same static configurations
that were found in section 3 are still static configurations in the above
background, thus showing their BPS character ~\cite{arkady}. 
Notice that this computation also gives further evidence for the physical 
interpretation given to the type IIA configuration obtained in the last 
subsection.

Using the same ansatz as in section 3, the non-trivial angular momenta
is given by
\begin{equation}
  P_\varphi = T_{\text{D}0}\frac{e^{-\phi}}{\sqrt{-\dot{x}^m\dot{x}^n
  g_{mn}}}\dot{x}^mg_{m\varphi} + T_{\text{D}0}\,C_\varphi
 \label{mom1}
\end{equation}
and
\begin{equation}
  P_\psi = T_{\text{D}0}\frac{e^{-\phi}}{\sqrt{-\dot{x}^m\dot{x}^n
  g_{mn}}}\dot{x}^mg_{m\psi} + T_{\text{D}0}\,C_\psi ~.
 \label{mom2}
\end{equation}

Requiring the configuration to be static 
$\left(P_\varphi =  P_\psi = 0\right)$, determines three
inequivalent solutions. It is life-reaffirming to check that these are 
precisely the ones found in our previous analysis : $r_0=0$, corresponding to a
D0-brane moving on the F5-brane, $\sin 2\theta_0=0$, corresponding to a 
D0-brane moving in the 67-plane or in the 89-plane and $\dot{\psi}=0$.
As in that case, the angular velocity in the last two solutions
is
\[
\omega_0+\cos 2\theta_0 v_0=-\beta~.
\]
Since, by construction, the configuration has vanishing momentum, its energy
equals minus the value of the lagrangian density evaluated on it. Again,
it is satisfactory to check that
\[
E=T_{\text{D}0}~.
\]
This result can be confirmed by solving, after some algebra, 
the mass shell condition for $E$
\begin{multline*}
  g^{00}\left(E+T_{\text{D}0}C_0\right)^2 + 2g^{0\varphi}\left(E+T_{\text{D}0}
  C_0\right)T_{\text{D}0}C_\varphi + 2g^{0\psi}\left(E+T_{\text{D}0}C_0\right)
  T_{\text{D}0}C_\psi \\
  + T_{\text{D}0}^2\left[g^{\varphi\varphi}C_\varphi^2 + g^{\psi\psi}C_\psi^2
  + 2 g^{\varphi\psi}C_\varphi C_\psi\right]^2 + T_{\text{D}0}^2\,U^{-3/2}
  \Delta^{-3/2} = 0~.
\end{multline*}

\section{Static D0-D2 bound state}

In a flat and topologically trivial spacetime, either D0-branes or D2-branes 
can be located anywhere,
the resulting configurations being stable and supersymmetric. Furthermore,
non-threshold D0-D2 bound states preserving one half of the spacetime
supersymmetry are known to exist. These can be realized on the D2-brane 
effective action as constant magnetic fluxes. We have learnt in the previous
section how to describe static, supersymmetric D0-branes in
F5-brane backgrounds. It is natural to ask whether the forementioned
bound states are also allowed in this new F5-brane backgrounds.

First of all, just as for the flat spacetime, one needs to identify the right
vacuum state describing a D2-brane. It is easy to show that a D2-brane
parallel to the F5-brane sitting a distance $r_0$ from it, feels no force.
In other words, the potential is constant and equals the tension of the brane.
It preserves one fourth of the spacetime supersymmetry.

It is natural to guess that the description of a D0-D2 bound state in a 
F5-brane background involves switching on some constant magnetic
flux on the brane, $F=2\pi\alpha^\prime\,F_{12}$, and giving some angular
velocity to the system $(\varphi=\omega\tau + \varphi_0)$,  such that the 
total angular momentum $P_\varphi$ vanishes, to compensate for the effect
of the non-trivial background geometry\footnote{We decided to concentrate
on this description for the D0-brane, but the other two
static configurations described earlier in this note would lead to exactly
the same picture found here.}. Using the static gauge, setting the 
rest of transverse coordinates to constant values and the electric components 
of the gauge field to a pure gauge configuration, the effective action
description the D2-brane at low energies and weak string coupling constant
is
\begin{equation}
  S=-T_{\text{D}2}\int d^{2+1}\sigma\,\Lambda_0^{-1}\,
  \left\{\sqrt{\Lambda_0 - (r\omega)^2}
  \sqrt{\Lambda_0 + F^2}- \beta r^2 F\omega\right\}\,.
\end{equation}
The physical requirement of vanishing angular momentum fixes the angular 
velocity to be
\[
  P_\varphi = \frac{\partial\cL}{\partial\omega} =0 \quad
  \Rightarrow \quad \omega_\star = -\frac{\beta F}{\sqrt{1+F^2}}\,,
\]
for which the energy density of the configuration equals
\[
E[\omega_\star]=T_{\text{D}2}\sqrt{1+F^2}\,,
\]
the energy density of a D0-D2 non-threshold bound state ~\cite{Polchinski}.
We conclude that such bound states exist in these new sectors
of string theory described by F5-branes.

In the same philosophy followed in the previous section, we could compute
the backreaction of the D0-D2 bound state, by taking the strong coupling
limit and considering the eleven dimensional supergravity description
of the above system. Locally, this should be given by an M2-brane boosted
in the eleventh compact dimension ~\cite{russo1}. By Kaluza--Klein reduction
along the orbits of the Killing vector $\xi_{\text{F5}}$, one would obtain
the desired type IIA configuration.

\section{Meta-stabilization of D2-branes}

The existence of a linear coupling among the angular velocity and the
RR 1-form was crutial for the stabilization mechanism of D0-branes in 
F5-brane backgrounds to work. D2-branes do also couple linearly
to such RR 1-form through the electric components of the gauge field
strength $F_{0a}$ $a=1,2$. Whenever the relative orientation between
the probe and the F5-brane is such that the above coupling is non-vanishing,
the electric field on the brane 
$(E^a = (2\pi\alpha^\prime)^{-1}\frac{\partial \cL}{\partial F_{0a}})$ 
will not vanish, even if $F_{0a}=0$. It will be examined below whether by 
requiring $E^a$ to vanish, the configuration is stabilized or not. 
As when discussing D0-branes in F5-brane backgrounds, it is useful to interpret
the corresponding configurations in M-theory. In such a strong coupling
limit, membranes propagate in a locally flat spacetime satisfying topologically
non-trivial conditions. Since the relation among the world-volume membrane
description and the world-volume D2-brane description is a world-volume
dualisation in 1+2 dimensions of the scalar field $x^\sharp(\sigma)$
parametrising the eleventh dimension ~\cite{four}
\begin{equation*}
  dx^\sharp + \hat{C}_{(1)} = \left(e^\Phi\,\sqrt{-\text{det}\,({\cal G} +
  \cF)}\right)\star\cF\,,
\end{equation*}
where $\cF=(2\pi\alpha^\prime)F + \hat{B}_{(2)}$, it is apparent that
we do need no non-trivial embedding $x^\sharp=x^\sharp(\sigma)$ in eleven
dimensions to describe the corresponding D2-brane configurations. Indeed,
the source for the non-vanishing electric components of the field strength
$F$ is entirely given in terms of the pull-back $\hat{C}_{(1)}$. We shall
then be interested in those cases where such a pull-back is non-vanishing,
the latter depending on the relative orientations between the probe worldspace
$\Sigma$ and the F5-brane background. In all of them, the eleven dimensional
configuration corresponds to a curved membrane in a locally flat spacetime.

\vspace{0.5cm}
\underline{$\Sigma=\Sigma_{\varphi\psi}$}
\vspace{0.5cm}

This corresponds to a D2-brane in the $\varphi\psi$-plane. When $F_{0a}=0$,
the potential felt by the probe 
\[
  V=T_{D2}r_0^2\sin 2\theta_0\left\{1+3\left(\beta r_0\right)^2\Lambda_0^{-1}
  \right\}^{1/2} ~,
\]
depends on the distance $r_0$ to the F5-brane and the $\theta_0$ angle
among the $67$-plane and $89$-plane. It will be useful to distinguish
between the first factor $r_0^2\sin 2\theta_0$ and the second 
$\left\{1+3\left(\beta r_0\right)^2\Lambda_0^{-1}\right\}^{1/2}$. The first
one accounts for the curved nature of the brane probe worldspace, whereas the
second one encodes the non-trivial background geometry and dilaton.
Notice that the above configuration collapses to a pointlike configuration
$(r_0\to 0)$.

Whenever $F_{0a}\neq 0$, the effective action is given by
\begin{multline}
  S = -T_{\text{D}2}\int d^{1+2}\sigma \,\Lambda_0^{-1}r_0
  \left\{\left[r_0^2\sin^2 2\theta_0 \Lambda_0 
  - (F_\psi-\cos 2\theta_0 F_\varphi)^2\right. \right. \\
  \left. \left. - \Lambda_0\sin^ 2 2\theta_0 F_\varphi^2\right]^{1/2}
  - \beta r_0 (F_\psi-\cos 2\theta_0 F_\varphi)\right\}~,
 \label{meta1}
\end{multline}
where $F_\psi = 2\pi\alpha^\prime\,F_{0\psi}$ and $F_\varphi = 
2\pi\alpha^\prime\,F_{0\varphi}$. The physical requirement of vanishing 
electric field on the brane determines the extremal values
\[
F^\star_\varphi = 0 \qquad , \qquad 
F^\star_\psi = -(\beta r_0)\,r_0\sin 2\theta_0 ~.
\]
When one computes the energy density $(\cE)$ for such configuration, it 
equals minus the value of the lagrangian density, since all momenta and 
electric field $E^a$ vanish. This energy density equals
\[
\cE[F^\star_\varphi\,,F^\star_\psi]= T_{\text{D}2}r_0^2\sin 2\theta_0 ~.
\]
Thus, by switching on $F^\star_\psi$ on the brane, the physical effects due 
to the curved geometry, non-vanishing dilaton and presence of RR 1-form
are screened. The energy of the configuration is the same as that for a curved
D2-brane in a flat background, thus confirming the eleven dimensional
interpretation of the configuration. As in that case, this configuration will 
not be stable since there is no force that prevents the D2-brane from 
contracting to a point $(r_0\to 0)$.

\vspace{0.5cm}
\underline{$\Sigma=\Sigma_{\theta\varphi}$}
\vspace{0.5cm}

This corresponds to a D2-brane in the $\theta\varphi$-plane. It will be 
checked that the same screening effect described before happens here. We shall
switch on $F=2\pi\alpha^\prime\,F_{0\theta}$ on the brane, such that the 
effective action is
\begin{equation}
  S= -T_{\text{D}2}\int d^{1+2}\sigma \Lambda_0^{-1} r_0
  \left\{\sqrt{r_0^2\Lambda_0 - F^2} - \beta r_0\,F\right\}~.
 \label{meta2}
\end{equation}
Requiring the electric field to vanish, fixes
\[
F_\star = -(\beta r_0)r_0~,
\]
for which the energy density of the configuration equals
\[
\cE[F_\star]= T_{\text{D}2}r_0^2\,,
\]
the one of a curved D2-brane in a flat spacetime, as expected from the eleven
dimensional discussion. As before, the configuration
would collapse to zero size, but all effects due to the original curved
background were removed by $F_\star$.

\vspace{0.5cm}
\underline{$\Sigma=\Sigma_{\theta\psi}$}
\vspace{0.5cm}

For completeness, we shall discuss a D2-brane on the $\theta\psi$-plane, even
though the conclusion is entirely analogous to the previous configurations.
The effective action describing the system when 
$F=2\pi\alpha^\prime F_{0\theta}$ is different from zero, is given by
\begin{equation}
  S = T_{\text{D}2}\Lambda_0^{-1} r_0
  \left\{\sqrt{1+(\beta r_0)^2\sin^2 2\theta_0}
  \sqrt{r_0^2\Lambda_0 - F^2} - \beta r_0\,F\cos 2\theta_0\right\}~,
 \label{meta3}
\end{equation}
and the requirement of vanishing electric field fixes
\[
F_\star = -(\beta r_0)r_0\cos 2\theta_0~,
\]
for which the energy density equals 
\[
\cE[F_\star]= T_{\text{D}2}r_0^2~,
\]
which can be interpreted as in previous discussions.

\medskip
\section*{Acknowledgments}
\noindent
The author would like to thank O. Aharony and M. Berkooz for pointing
out a mistake in the first version of this note and O. Aharony for proof
reading this manuscript. This research has been supported by a Marie Curie 
Fellowship of the European Community programme ``Improving the Human Research 
Potential and the Socio-Economic knowledge Base'' under the contract number 
HPMF-CT-2000-00480.


\appendix

\section{Supersymmetry}

The type IIA F5-brane configuration is obtained from Kaluza-Klein
reduction of the flat eleven dimensional Minkowski space along the orbits
generated by the Killing vector
\begin{equation}
  \xi_{\text{F5}} = \partial_{x^\sharp} 
  + \beta\left(x^6\6_7-x^7\6_6 + x^8\6_9 - x^9\6_8\right)~.
 \label{vackill}
\end{equation}
The Killing vector $\xi_{\text{F5}}$ acts on the constant Killing spinor 
$\varepsilon$ via the spinorial Lie derivative (see, e.g., ~\cite{Kosmann} 
and also ~\cite{JMFKilling}), which in local coordinates is given by
\begin{equation}
  L_{\xi_{\text{F5}}} \varepsilon = \left((\xi_{\text{F5}})^m\nabla_m 
  + \frac{1}{4}\6_{[m}(\xi_{\text{F5}})_{n]}
  \Gamma^{mn} \right)\varepsilon
\end{equation}
The condition that $\xi$ preserves some supersymmetry $(L_\xi\varepsilon=0)$
is equivalent to
\[
\Gamma_{6789}\,\varepsilon=\varepsilon~,
\]
where it was assumed that $\beta\neq 0$. This corresponds to having a F5-brane
in the 12345-plane. Thus, F5-branes preserve one half of the spacetime
supersymmetry.

The above computation was done in eleven dimensions in a coordinate
system in which the eleven dimensional geometry is manifestly flat.
When dealing with kappa symmetry and supersymmetry in type IIA/B string
theories, one requires the explicit form for the corresponding Killing
spinors associated with a given bosonic supersymmetric background. 
In the case of F5-branes, the dilatino and gravitino supersymmetry
transformations reduce to
\begin{eqnarray}
  \delta\lambda &=& \G^m\partial_m\Phi\varepsilon + \frac{3}{8}e^\Phi
  \G^{mn}F_{mn}\G_\sharp\varepsilon \nonumber \\
  \delta\psi_m &=& \left(\partial_m +\frac{1}{4}\omega_m~^{ab}
  \G_{\underline{ab}}\right)\varepsilon + \frac{1}{16}e^\Phi \G^{pq}F_{pq}
  \G_m \G_\sharp\varepsilon \nonumber ~.
\end{eqnarray}
Thus, obtaining the Killing spinors is equivalent to solving $\delta\lambda =
\delta\psi_m =0$, when the background is the F5-brane one.

It is useful to introduce an orthonormal basis
\begin{eqnarray}
  e^{\underline{\mu}} &=& \Lambda^{1/4}\,dx^\mu \quad \mu=0,\dots ,5 
  \nonumber \\
  e^{\underline{r}} &=& \Lambda^{1/4}\,dr \quad , \quad 
  e^{\underline{\theta}} = \Lambda^{1/4}\,r\,d\theta \nonumber \\
  e^{\underline{\psi}} &=& \Lambda^{1/4}\,r\sin 2\theta d\psi \quad , \quad 
  e^{\underline{\varphi}} = \Lambda^{-1/4}\,r\left(d\varphi + \cos 2\theta
  d\psi\right)~. \nonumber
\end{eqnarray} 
In such a basis, the RR 2-form field strength $F_2=dA_1$ is given by
\begin{equation*}
  F_2 = 2\beta(\Lambda^{-2}\,e^{\underline{r}}\wedge
  e^{\underline{\varphi}} - \Lambda^{-3/2}e^{\underline{\theta}}\wedge
  e^{\underline{\psi}})~,
\end{equation*}
whereas the non-trivial spin connection reads as
\begin{eqnarray}
  \omega^{\underline{\mu}}~_{\underline{r}} &=& \frac{1}{2}(\beta r)r
  \Lambda^{-5/4}\, e^{\underline{\mu}} \quad , \quad
  \omega^{\underline{\psi}}~_{\underline{r}} = r^{-1}\Lambda^{-5/4}
  (\Lambda + \frac{1}{2}(\beta r)^2) e^{\underline{\psi}} 
  \nonumber \\
  \omega^{\underline{\theta}}~_{\underline{r}} &=& r^{-1}\Lambda^{-5/4}
  (\Lambda + \frac{1}{2}(\beta r)^2) e^{\underline{\theta}}
  \quad , \quad
  \omega^{\underline{\varphi}}~_{\underline{r}} = r^{-1}\Lambda^{-5/4}
  (\Lambda - \frac{1}{2}(\beta r)^2) e^{\underline{\varphi}} 
  \nonumber \\
  \omega^{\underline{\psi}}~_{\underline{\theta}} &=& r^{-1}\Lambda^{-3/4}
  e^{\underline{\varphi}} + 2r^{-1}\Lambda^{-1/4}
  \frac{\cos 2\theta}{\sin 2\theta} e^{\underline{\psi}} \nonumber \\
  \omega^{\underline{\psi}}~_{\underline{\varphi}} &=& r^{-1}\Lambda^{-3/4}
  e^{\underline{\theta}} \quad , \quad
  \omega^{\underline{\varphi}}~_{\underline{\theta}} = r^{-1}\Lambda^{-3/4}
  e^{\underline{\psi}}~. \nonumber
\end{eqnarray}

The dilatino equation gives rise to an algebraic equation
\[
\left\{\beta r \G_{\underline{r}} - \Lambda^{1/2}\G_{\underline{\theta\psi}}
\G_\sharp + \G_{\underline{r\varphi}}\G_\sharp\right\}\varepsilon = 0~,
\]
this being equivalent to
\begin{equation}
  \left\{\beta r\G_{\underline{\varphi}}\G_\sharp - \Lambda^{1/2}
  \G_{\underline{r\theta\psi\varphi}}\right\}\varepsilon = \varepsilon~.
 \label{dilatino}
\end{equation}
On the other hand, if we consider the linear combinations 
$e_{\underline{a}}~^m\delta\psi_m = 0$, the gravitino equation can be 
rewritten as
\begin{multline}
  \left(e_{\underline{a}}~^m\partial_m + \frac{1}{4}
  \omega_{\underline{a}}~^{\underline{bc}}\G_{\underline{bc}}\right)\varepsilon
  -\frac{1}{4}\Lambda^{-5/4}\beta^2 r \G_{\underline{a}}\G_{\underline{r}}
  \varepsilon \\
  -\frac{1}{4}\Lambda^{3/4}\delta_a~^{[b}\G^{\underline{c}]}F_{\underline{bc}}
  \G_\sharp\varepsilon = 0~,
 \label{gravitino1}
\end{multline}
where we have already used \eqref{dilatino}. Notice that \eqref{gravitino1}
is trivially satisfied for $\underline{a}=\underline{\mu}$, due to
\eqref{dilatino}. The remaining equations 
\begin{equation}
  \begin{aligned}[m]
    \partial_r\varepsilon &= \frac{1}{2}\Lambda^{-1}\beta(\frac{1}{2}
    \beta r + \G_{\underline{\varphi}}\G_\sharp)\varepsilon \\
    \partial_\theta \varepsilon &= -\frac{1}{2}\left[\G_{\underline{\theta r}}
    + \Lambda^{-1/2}\G_{\underline{\theta}}(\G_{\underline{\varphi}} +
    \beta r \G_\sharp)\right]\varepsilon \\
    (\partial_\psi -\cos 2\theta\partial_\varphi )\varepsilon &= 
    -\frac{1}{2}\sin 2\theta\left[\G_{\underline{\psi r}} + \Lambda^{-1/2}
    \G_{\underline{\theta}}(\G_{\underline{\varphi}} 
    + \beta r \G_\sharp)\right]\varepsilon  \\
    & - \cos 2\theta \G_{\underline{\psi\theta}}\varepsilon \\
    \partial_\varphi\varepsilon &= -\frac{1}{2}\Lambda^{-1}\left[
    \Lambda^{-1/2}\G_{\underline{\varphi r}} + \G_{\underline{\varphi\theta}}
    + \Lambda^{-1/2}\beta r\G_{\underline{r}}\G_\sharp\right]
  \end{aligned}
\end{equation}
suggest the form of the Killing spinors should be
\[
  \varepsilon = M[r,\dots,\psi]\Pi_l\,e^{f_l(r,\dots,\psi)
  \G^{(l)}}\varepsilon_0~,
\]
where $\varepsilon_0$ is some constrained constant spinor and $\G^{(l)}$
stands for appropiate antisymmetrised products of gamma matrices,
whereas $f_l(r,\dots,\psi)$ are generically non-constant functions
just as $M[r,\dots,\psi]$, with the difference that the latter might
also be a matrix.


\end{document}